\documentclass[showpacs,twocolumn,floatfix,superscriptaddress]{revtex4}
\usepackage{graphicx}
\usepackage{amsmath}
\begin{document}
\title{Temperature dependent third cumulant of tunneling noise}
\author{C. W. J. Beenakker}
\affiliation{Instituut-Lorentz, Universiteit Leiden, P.O. Box 9506, 2300 RA
Leiden, The Netherlands}
\author{M. Kindermann}
\affiliation{Instituut-Lorentz, Universiteit Leiden, P.O. Box 9506, 2300 RA
Leiden, The Netherlands}
\author{Yu.\ V. Nazarov}
\affiliation{Department of Nanoscience, Delft University of Technology,
Lorentzweg 1, 2628 CJ Delft, The Netherlands}
\date{January 2003}
\begin{abstract}
Poisson statistics predicts that the shot noise in a tunnel junction has a
temperature independent third cumulant $e^{2}\bar{I}$, determined solely by the
mean current $\bar{I}$. Experimental data, however, show a puzzling temperature
dependence. We demonstrate theoretically that the third cumulant becomes
strongly temperature dependent and may even change sign as a result of feedback
from the electromagnetic environment. In the limit of a noninvasive
(zero-impedance) measurement circuit in thermal equilibrium with the junction,
we find that the third cumulant crosses over from $e^{2}\bar{I}$ at low
temperatures to $-e^{2}\bar{I}$ at high temperatures.
\end{abstract}
\maketitle

Shot noise of the electrical current was studied a century ago as a way to
measure the fundamental unit of charge \cite{Sch18}. Today shot noise is used
for this purpose in a wide range of contexts, including superconductivity and
the fractional quantum Hall effect \cite{Bla00}. Already in the earliest work
on vacuum tubes it was realized that thermal fluctuations of the current can
mask the fluctuations due to the discreteness of the charge. In semiconductors,
in particular, accurate measurements of shot noise are notoriously difficult
because of the requirement to maintain a low temperature at a high applied
voltage.

Until very recently, only the second cumulant of the fluctuating current was
ever measured. The distribution of transferred charge is nearly Gaussian,
because of the law of large numbers, so it is quite nontrivial to extract
cumulants higher than the second.  Much of the experimental effort was
motivated by Levitov and Reznikov's prediction \cite{Lev01} that odd cumulants
of the current through a tunnel junction should not be affected by the thermal
noise that contaminates the even cumulants. This is a direct consequence of the
Poisson statistics of tunneling events. The third cumulant should thus have the
linear dependence on the applied voltage characteristic of shot noise,
regardless of the ratio of voltage and temperature. In contrast, the second
cumulant levels off at the thermal noise for low voltages.

The first experiments on the voltage dependence of the third cumulant of tunnel
noise have now been reported \cite{Reu02}. The pictures are strikingly
different from what was expected theoretically. The slope varies by an order of
magnitude between low and high voltages, and for certain samples even changes
sign. Such a behavior is expected for a diffusive conductor \cite{Gut02}, but
not for a tunnel junction. Although the data is still preliminary, it seems
clear that an input of new physics is required for an understanding. It is the
purpose of this paper to provide such input.

We will show that the third cumulant of the measured noise (unlike the second
cumulant \cite{Gav02}) is affected by the measurement circuit in a nonlinear
way. The effect can be seen as a backaction of the electromagnetic environment
\cite{Kin03}. We have found that the backaction persists even in the limit of
zero impedance, when the measurement is supposed to be noninvasive. The
temperature independent result for the third cumulant of tunneling noise is
recovered only if the measurement circuit has both negligible impedance and
negligible temperature.

\begin{figure}
\includegraphics[width=6cm]{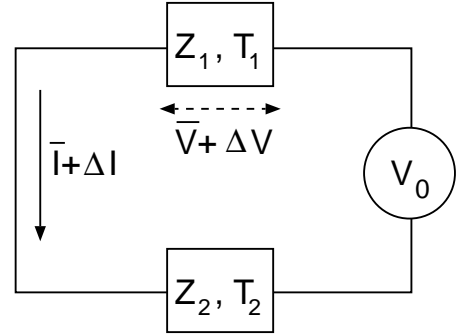}
\caption{
Two resistors in series with a voltage source. The fluctuating current and
voltage are indicated.
\label{circuit2}
}
\end{figure}

The circuit is shown schematically in Fig.\ \ref{circuit2}. Two resistors
(impedances $Z_{1}$, $Z_{2}$ and temperatures $T_{1}$, $T_{2}$) are connected
in series to a voltage source (voltage $V_{0}$). We will specialize later to
the case that resistor 1 is a tunnel junction and that resistor 2 represents
the macroscopic measurement circuit, but our main results hold for any two
resistors. We disregard possible Coulomb blockade effects on fluctuations
\cite{Ben83,Lee96,Gal02}, which is justified if the impedances at frequencies
of order $eV/\hbar$ are small compared to $\hbar/e^{2}$ \cite{ASI}.

We have calculated the temperature dependence of the third cumulant by two
altogether different methods, the Keldysh formalism \cite{Naz99} and the
Langevin approach \cite{Nag02}. The equivalence of the two methods has already
been demonstrated for a single resistor in the absence of any measurement
circuit \cite{Gut02}. Likewise, we have obtained the same results in both
calculations of the backaction from the measurement. We choose to present the
Langevin approach in this paper, because it can be explained in elementary
terms and provides an intuitive physical insight.

Starting point of the Langevin approach is the separation of the fluctuation
$\Delta I_{i}$ of the current through resistor $i=1,2$ into an intrinsic
fluctuation $\delta I_{i}$ plus a term induced by a fluctuation $\Delta V_{i}$
of the voltage over the resistor: $\Delta I_{i}=\delta I_{i}+\Delta
V_{i}/Z_{i}$. At low frequencies $\Delta I_{1}=\Delta I_{2}\equiv\Delta I$ and
$\Delta V_{1}=-\Delta V_{2}\equiv\Delta V$. Upon substitution we arrive at the
two equations
\begin{equation}
Z\Delta I=Z_{1}\delta I_{1}+Z_{2}\delta I_{2},\;\;Z\Delta V=Z_{1}Z_{2}(\delta
I_{2}-\delta I_{1}),\label{Langevin}
\end{equation}
where $Z=Z_{1}+Z_{2}$ is the total impedance of the circuit.

For simplicity we assume that $Z_{i}$ is real and frequency independent in the
frequency range of the measurement. All formulas have a straightforward
generalization to complex $Z_{i}(\omega)$. We do not need to assume at this
stage that the current-voltage characteristic of the resistors is linear. If it
is not, then one should simply replace $1/Z_{i}$ by the differential
conductance evaluated at the mean voltage $\overline{V_{i}}$ over the resistor.

The mean voltages are given by
$\overline{V_{1}}=(Z_{1}/Z)V_{0}\equiv\overline{V}$ and
$\overline{V_{2}}=V_{0}-\overline{V}$. The intrinsic current fluctuations
$\delta I_{i}$ are driven by the fluctuating voltage
$V_{i}=\overline{V_{i}}+\Delta V_{i}$, and therefore depend in a nonlinear way
on $\Delta V$. The nonlinearity has the effect of mixing in lower order
cumulants of $\delta I_{i}$ in the calculation of the $p$-th cumulant of
$\Delta I$, starting from $p=3$.

Before addressing the case $p=3$ we first consider $p=2$, when all averages
$\langle\cdots\rangle_{\overline{V}}$ can be performed at the mean voltage. At
low frequencies one has
\begin{equation}
\langle\delta I_{i}(\omega)\delta
I_{i}(\omega')\rangle_{\overline{V}}=2\pi\delta(\omega+\omega'){\cal
C}_{i}^{(2)}(\overline{V_{i}}).\label{S2def}
\end{equation}
The noise power ${\cal C}_{i}^{(2)}$ depends on the model for the resistor. We
give two examples. In a macroscopic resistor the shot noise is suppressed by
electron-phonon scattering and only thermal noise remains:
\begin{equation}
{\cal C}_{i}^{(2)}=2kT_{i}/Z_{i}\label{Sthermal}
\end{equation}
at temperature $T_{i}$, independent of the voltage. (The noise power is a
factor of two larger if positive and negative frequencies are identified.) In a
tunnel junction both thermal noise and shot noise coexist, according to
\cite{Bla00}
\begin{equation}
{\cal C}_{i}^{(2)}(\overline{V}_{i})=(e\overline{V}_{i}/Z_{i})\,{\rm
cotanh}\,(e\overline{V}_{i}/2kT_{i}).\label{S2V}
\end{equation}

{}From Eq.\ (\ref{Langevin}) we compute the correlator
\begin{equation}
\langle\Delta X(\omega)\Delta
Y(\omega')\rangle_{\overline{V}}=2\pi\delta(\omega+\omega')
S_{XY}(\overline{V}),\label{SXYdef}
\end{equation}
where $X$ and $Y$ can represent $I$ or $V$. The result is
\begin{subequations}
\label{SXYresult}
\begin{eqnarray}
S_{II}&=&Z^{-2}[Z_{1}^{2}{\cal C}_{1}^{(2)}(\overline{V})+Z_{2}^{2}{\cal
C}_{2}^{(2)}(V_{0}-\overline{V})],\label{SIIresult}\\
S_{VV}&=&Z^{-2}(Z_{1}Z_{2})^{2}[{\cal C}_{1}^{(2)}(\overline{V})+{\cal
C}_{2}^{(2)}(V_{0}-\overline{V})],\label{SVVresult}\\
S_{IV}&=&Z^{-2}Z_{1}Z_{2}[Z_{2}{\cal
C}_{2}^{(2)}(V_{0}-\overline{V})-Z_{1}{\cal
C}_{1}^{(2)}(\overline{V})].\label{SIVresult}
\end{eqnarray}
\end{subequations}

Eq.\ (\ref{SXYdef}) applies to a time independent mean voltage $\overline{V}$.
For a time dependent perturbation $v(t)$ one has, to linear order,
\begin{eqnarray}
\langle\Delta X(\omega)\Delta Y(\omega')\rangle_{\overline{V}+v}=\langle\Delta
X(\omega)\Delta Y(\omega')\rangle_{\overline{V}}\nonumber\\
\mbox{}+v(\omega+\omega')\frac{d}{d\overline{V}}S_{XY}(\overline{V}).
\label{SXYv}
\end{eqnarray}
We will use this equation, with $v=\Delta V$, to describe the effect of a
fluctuating voltage over the resistors. This assumes a separation of time
scales between $\Delta V$ and the intrinsic current fluctuations $\delta
I_{i}$, so that we can first average over $\delta I_{i}$ for given $\Delta V$
and then average over $\Delta V$.

Turning now to the third cumulant, we first note that at fixed voltage the
intrinsic current fluctuations $\delta I_{1}$ and $\delta I_{2}$ are
uncorrelated, with third moment
\begin{equation}
\langle\delta I_{i}(\omega_{1})\delta I_{i}(\omega_{2})\delta
I_{i}(\omega_{3})\rangle_{\overline{V}}=2\pi\delta(\omega_{1}+\omega_{2}+\omega_{3})
{\cal C}_{i}^{(3)}(\overline{V_{i}}).\label{S3def}
\end{equation}
The spectral density ${\cal C}_{i}^{(3)}$ vanishes for a macroscopic resistor.
For a tunnel junction it has the temperature independent value \cite{Lev01}
\begin{equation}
{\cal
C}_{i}^{(3)}(\overline{V_{i}})=e^{2}\overline{V_{i}}/Z_{i}=
e^{2}\bar{I},\label{S3V}
\end{equation}
with $\bar{I}$ the mean current.

We introduce the nonlinear feedback from the voltage fluctuations through the
relation
\begin{eqnarray}
&&\langle\Delta X_{1}\Delta X_{2}\Delta X_{3}\rangle=\langle\Delta X_{1}\Delta
X_{2}\Delta X_{3}\rangle_{\overline{V}}\nonumber\\
&&\;\;\mbox{}+\sum_{\rm cyclic}\langle\Delta X_{j}\Delta
V(\omega_{k}+\omega_{l})\rangle_{\overline{V}}\frac{d}{d\overline{V}}
S_{X_{k}X_{l}}(\overline{V}).\label{IIInonlinear}
\end{eqnarray}
The variable $X_{j}$ stands for $I(\omega_{j})$ or $V(\omega_{j})$ and the sum
is over the three cyclic permutations $j,k,l$ of the indices $1,2,3$. These
three terms account for the fact that the same voltage fluctuation $\Delta V$
that affects $S_{X_{k}X_{l}}$ also correlates with $X_{j}$, resulting in a
cross-correlation.

Eq.\ (\ref{IIInonlinear}) has the same form as the ``cascaded average'' through
which Nagaev introduced a nonlinear feedback into the Langevin equation
\cite{Nag02}. In that work the nonlinearity appears because the Langevin source
depends on the electron density, which is itself a fluctuating quantity --- but
on a slower time scale, so the averages can be carried out separately, or
``cascaded''. In our case the voltage drop $\Delta V_{i}$ over the resistors is
the slow variable, relative to the intrinsic current fluctuations $\delta
I_{i}$.

Eq.\ (\ref{IIInonlinear}) determines the current  and voltage correlators
\begin{equation}
\langle\Delta X(\omega_{1})\Delta Y(\omega_{2})\Delta
Z(\omega_{3})\rangle=2\pi\delta(\omega_{1}+\omega_{2}+\omega_{3})
C_{XYZ}(\overline{V}). \label{SXYZdef}
\end{equation}
We find
\begin{widetext}
\begin{subequations}
\label{Cresult}
\begin{eqnarray}
C_{III}&=&Z^{-3}[Z_{1}^{3}{\cal C}_{1}^{(3)}(\overline{V})+Z_{2}^{3}{\cal
C}_{2}^{(3)}(V_{0}-\overline{V})]+3S_{IV} \frac{d}{d\overline{V}}S_{II},
\label{CIresult}\\
C_{VVV}&=&Z^{-3}(Z_{1}Z_{2})^{3}[{\cal C}_{2}^{(3)}(V_{0}-\overline{V})-{\cal
C}_{1}^{(3)}(\overline{V})]+3S_{VV}
\frac{d}{d\overline{V}}S_{VV},\label{CVresult}\\
C_{VVI}&=&Z^{-3}(Z_{1}Z_{2})^{2}[Z_{1}{\cal
C}_{1}^{(3)}(\overline{V})+Z_{2}{\cal C}_{2}^{(3)}(V_{0}-\overline{V})]+2S_{VV}
\frac{d}{d\overline{V}}S_{IV}+S_{IV}
\frac{d}{d\overline{V}}S_{VV},\label{CVVIresult}\\
C_{IIV}&=&Z^{-3}Z_{1}Z_{2}[Z_{2}^{2}{\cal
C}_{2}^{(3)}(V_{0}-\overline{V})-Z_{1}^{2}{\cal C}_{1}^{(3)}(\overline{V})]+
2S_{IV}\frac{d}{d\overline{V}}S_{IV}+S_{VV}
\frac{d}{d\overline{V}}S_{II}.\label{CIIVresult}
\end{eqnarray}
\end{subequations}
\end{widetext}

We apply the general result (\ref{Cresult}) to a tunnel barrier (resistor
number 1) in series with a macroscopic resistor (number 2). The spectral
densities ${\cal C}_{1}^{(2)}$ and ${\cal C}_{1}^{(3)}$ are given by Eqs.\
(\ref{S2V}) and (\ref{S3V}), respectively. For ${\cal C}_{2}^{(2)}$ we use Eq.\
(\ref{Sthermal}), while ${\cal C}_{2}^{(3)}=0$. From this point on we assume
linear current-voltage characteristics, so $\overline{V}$-independent
$Z_{i}$'s. We compare $C_{I}\equiv C_{III}$ with $C_{V}\equiv
-C_{VVV}/Z_{2}^{3}$. The choice of $C_{V}$ is motivated by the typical
experimental situation in which one measures the current fluctuations
indirectly through the voltage over a macroscopic series resistor. From Eq.\
(\ref{Cresult}) we find
\begin{eqnarray}
C_{x}&=&\frac{e^{2}\bar{I}}{(1+Z_{2}/Z_{1})^{3}}\left[1+\frac{3(\sinh u\cosh
u-u)}{(1+Z_{1}/Z_{2})\sinh^{2}u}\right.\nonumber\\
&&\left.\mbox{}\times\left(\frac{T_{2}}{T_{1}}\frac{g_{x}}{u}-{\rm
cotanh}\,u\right)\right],\label{Cxresult}
\end{eqnarray}
with $g_{I}=1$, $g_{V}=-Z_{1}/Z_{2}$, and $u=e\overline{V}/2kT_{1}$.

In the shot noise limit ($e\overline{V}\gg kT_{1}$) we recover the third
cumulant obtained in Ref.\ \cite{Kin03} by the Keldysh technique:
\begin{equation}
C_{I}=C_{V}=e^{2}\bar{I}\frac{1-2Z_{2}/Z_{1}}{(1+Z_{2}/Z_{1})^{4}}.
\label{ClowT}
\end{equation}
In the opposite limit of small voltages ($e\overline{V}\ll kT_{1}$) we obtain
\begin{eqnarray}
C_{I}&=&e^{2}\bar{I}\frac{1+(Z_{2}/Z_{1})(2T_{2}/T_{1}-1)}
{(1+Z_{2}/Z_{1})^{4}},\label{CIhighT}\\
C_{V}&=&e^{2}\bar{I}\frac{1-Z_{2}/Z_{1}-2T_{2}/T_{1}}
{(1+Z_{2}/Z_{1})^{4}}.\label{CVhighT}
\end{eqnarray}

\begin{figure}
\includegraphics[width=6cm]{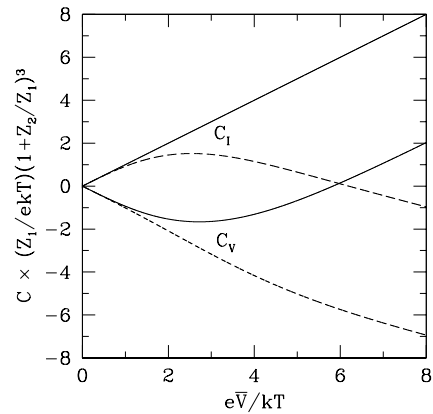}
\caption{
Voltage dependence of the third cumulants $C_{I}$ and $C_{V}$ of current and
voltage for a tunnel junction (resistance $Z_{1}$) in series with a macroscopic
resistor $Z_{2}$. The two solid curves are for $Z_{2}/Z_{1}\rightarrow 0$ and
the dashed curves for $Z_{2}/Z_{1}=1$. The curves are computed from Eq.\
(\protect\ref{Cxresult}) for $T_{1}=T_{2}\equiv T$. The high voltage slopes are
the same for $C_{I}$ and $C_{V}$, while the low voltage slopes have the
opposite sign.
\label{Cplot}
}
\end{figure}

We conclude that there is a change in the slope $dC_{x}/d\bar{I}$ from low to
high voltages.  If the entire system is in thermal equilibrium ($T_{2}=T_{1}$),
then the change in slope is a factor $\pm(Z_{1}-2Z_{2})(Z_{1}+Z_{2})^{-1}$,
where the $+$ sign is for $C_{I}$ and the $-$ sign for $C_{V}$. In Fig.\
\ref{Cplot} we plot the entire voltage dependence of the third cumulants.

The limit $Z_{2}/Z_{1}\rightarrow 0$ of a noninvasive measurement is of
particular interest. Then $C_{I}=e^{2}\bar{I}$ has the expected result for an
isolated tunnel junction \cite{Lev01}, but $C_{V}$ remains affected by the
measurement circuit:
\begin{equation}
\lim_{Z_{2}/Z_{1}\rightarrow
0}C_{V}=e^{2}\bar{I}\left(1-\frac{T_{2}}{T_{1}}\frac{3(\sinh u\cosh
u-u)}{u\sinh^{2}u}\right).\label{CVZ0limit}
\end{equation}
This limit is also plotted in Fig.\ \ref{Cplot}, for the case
$T_{2}=T_{1}\equiv T$ of thermal equilibrium between the tunnel junction and
the macroscopic series resistor. The slope then changes from
$dC_{V}/d\bar{I}=-e^{2}$ at low voltages to $dC_{V}/d\bar{I}=e^{2}$ at high
voltages. The minimum $C_{V}=-1.7\,ekT/Z_{1}=-0.6\,e^{2}\bar{I}$ is reached at
$e\overline{V}=2.7\,kT$.

In conclusion, we have demonstrated that feedback from the measurement circuit
introduces a temperature dependence of the third cumulant of tunneling noise.
The temperature independent result $e^{2}\bar{I}$ of an isolated tunnel
junction \cite{Lev01} acquires a striking temperature dependence in an
electromagnetic environment, to the extent that the third cumulant may even
change its sign. Precise predictions have been made for the dependence of the
noise on the environmental impedance and temperature, which can be tested in
ongoing experiments \cite{Reu02}.

We gratefully acknowledge discussions with B. Reulet, which motivated us to
write this paper. Our research is supported by the Dutch Science Foundation
NWO/FOM.

\end{document}